\documentclass{article}
\usepackage{preprint,times}

\usepackage[colorlinks=true,linkcolor=blue,citecolor=blue,urlcolor=blue]{hyperref}
\usepackage{url}
\usepackage{graphicx}
\usepackage{subcaption}
\usepackage{amsthm}
\usepackage{amsmath}
\usepackage{amssymb}
\usepackage{mathtools}
\usepackage{pifont}
\usepackage{booktabs}
\usepackage{arydshln}
\usepackage{xspace}
\usepackage{algorithm}
\usepackage{algorithmicx}
\usepackage{algpseudocode}
\usepackage{multirow}
\usepackage{bbm}
\usepackage{latexsym}
\usepackage[T1]{fontenc}
\usepackage[utf8]{inputenc}
\usepackage{microtype}
\usepackage{inconsolata}
\usepackage{wrapfig}
\usepackage{tcolorbox}
\tcbuselibrary{breakable}
\usepackage{xcolor}

\makeatletter
\DeclareRobustCommand\onedot{\futurelet\@let@token\@onedot}
\def\@onedot{\ifx\@let@token.\else.\null\fi\xspace}

\def\eg{\emph{e.g}\onedot}

\DeclareRobustCommand{\ourapproach}{{\sc Think-Anywhere}\xspace}
\DeclareRobustCommand{\ourapproachstar}{{\sc Think-Anywhere*}\xspace}
\DeclareRobustCommand{\ourapproachbf}{{\sc \textbf{Think-Anywhere}}\xspace}
\makeatother

\theoremstyle{definition}

\theoremstyle{remark}

\usepackage{marvosym} 

\newtcolorbox[use counter=prompt]{promptbox}[2][]{%
  colback=gray!5,
  colframe=black!50,
  fonttitle=\bfseries,
  title=Prompt: #2,
  breakable,
  #1
}

\iclrfinalcopy
\title{Think Anywhere in Code Generation}

\author{Xue Jiang$^{1,2,}$\textsuperscript{\Letter}, Tianyu Zhang$^{1}$, Ge Li$^{1,}$\textsuperscript{\Letter}, Mengyang Liu$^{1}$, Taozhi Chen$^{1}$, Zhenhua Xu$^{1}$, \\
\textbf{Binhua Li$^{2}$, Wenpin Jiao$^{1}$, Zhi Jin$^{1}$, Yongbin Li$^{2}$, Yihong Dong$^{1,2,}$\textsuperscript{\Letter}} \\
$^1$ School of Computer Science, Peking University \\
$^2$ Tongyi Lab, Alibaba Group \\
\texttt{\{jiangxue, dongyh\}@stu.pku.edu.cn} \quad \texttt{lige@pku.edu.cn}\\\
}

\begin{document}

\maketitle

\begin{abstract}
Recent advances in reasoning Large Language Models (LLMs) have primarily relied on upfront thinking, where reasoning occurs before final answer. However, this approach suffers from critical limitations in code generation, where upfront thinking is often insufficient as problems' full complexity only reveals itself during code implementation. Moreover, it cannot adaptively allocate reasoning effort throughout the code generation process where difficulty varies significantly. In this paper, we propose \ourapproach, a novel reasoning mechanism that enables LLMs to invoke thinking on-demand at any token position during code generation. We achieve \ourapproach by first teaching LLMs to imitate the reasoning patterns through cold-start training, then leveraging outcome-based RL rewards to drive the model's autonomous exploration of when and where to invoke reasoning.
Extensive experiments on four mainstream code generation benchmarks (i.e., LeetCode, LiveCodeBench, HumanEval, and MBPP) show that \ourapproach achieves state-of-the-art performance over both existing reasoning methods and recent post-training approaches, while demonstrating consistent generalization across diverse LLMs.
Our analysis further reveals that \ourapproach enables the model to adaptively invoke reasoning at high-entropy positions, providing enhanced interpretability.
\end{abstract}
\renewcommand{\thefootnote}{\fnsymbol{footnote}}

\footnotetext[1]{Work done during Xue Jiang and Yihong Dong's internship at Tongyi Lab.}
\footnotetext[2]{Our source code and data are available at \url{https://github.com/jiangxxxue/Think-Anywhere}.}

\renewcommand{\thefootnote}{\arabic{footnote}}

\section{Introduction} \label{sec:intro}

Recent advances in Large Language Models (LLMs) have demonstrated remarkable capabilities in code generation tasks~\citep{codellama,lozhkov2024starcoder2stackv2,guo2024deepseek,DBLP:journals/tosem/DongJJL24,DBLP:journals/corr/abs-2508-00083}. A pivotal breakthrough in this domain has been the integration of reasoning mechanisms, particularly exemplified by Chain-of-Thought (CoT) prompting~\citep{wei2022chain,jiang2024selfplanning}. Recent reasoning-optimized LLMs, such as industry-leading OpenAI's o1~\citep{jaech2024o1systemcard}, DeepSeek-R1~\citep{guo2025deepseek}, and Kimi K2~\citep{Kimik2}, have achieved unprecedented performance by scaling up reasoning through reinforcement learning (RL). These models are trained to first complete global planning and logical deliberation within an internal thinking block, and then proceed to generate the final output. This upfront thinking approach has become the dominant technical pathway for enhancing complex reasoning capabilities in code generation~\citep{jaech2024o1systemcard,DBLP:journals/corr/abs-2601-13240,guo2025deepseek}.

\begin{figure*}[t!]
    \centering
    \includegraphics[width=0.87\textwidth]{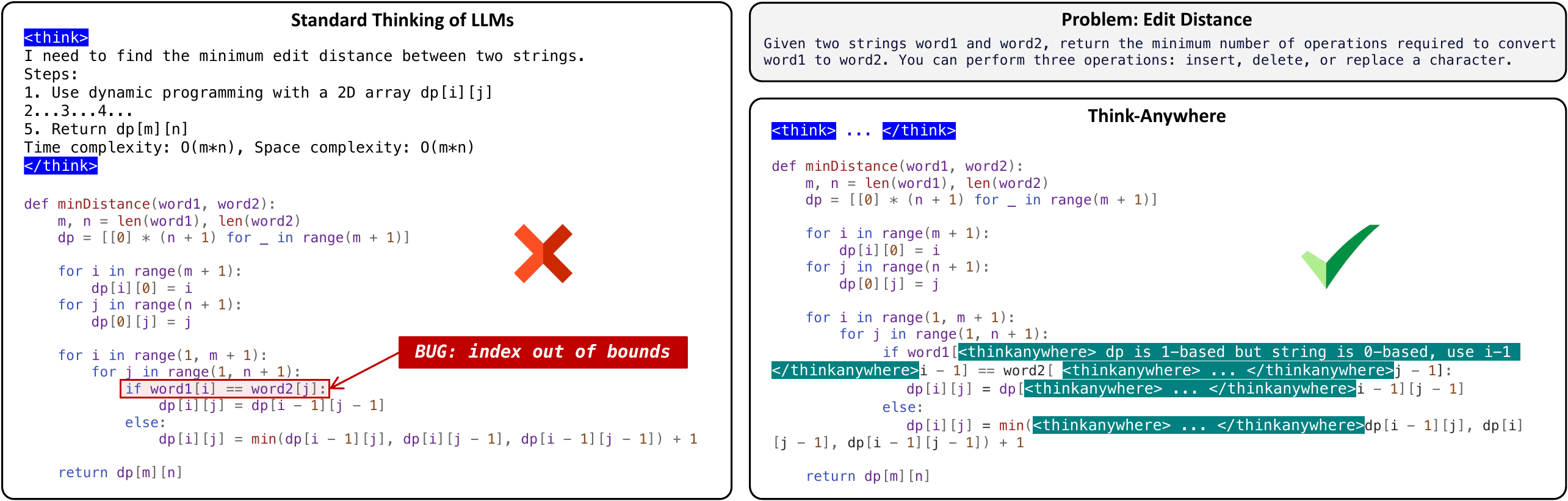}
    \caption{Illustration of \ourapproach. Reasoning can be invoked at any token position during code generation. The ellipsis (``...'') within <think> or <thinkanywhere> represents truncated thinking content for brevity.}
    \label{fig:intro_case}
\end{figure*}

While the upfront thinking approach has proven effective, it exhibits two limitations in code generation. First, upfront thinking is often insufficient, as the full complexity of problems typically only reveals itself during implementation. For instance, LLMs usually perform only plan-level thinking in the upfront reasoning phase, while new problems emerge during the code implementation stage, leading to bugs due to the lack of adequate reasoning, as shown in Figure \ref{fig:intro_case}. Second, upfront thinking cannot precisely allocate reasoning effort to the positions where it is needed. Different positions in code generation vary in difficulty, with simple boilerplate code requiring minimal computation while complex algorithmic decisions or edge case handling demanding deep reasoning.
By contrast, human coding cognition shows that developers not only think before coding but also pause to think at any point during implementation, which proves a more reasonable thinking approach.
Motivated by these observations, we desire a mechanism that enables models to invoke reasoning at any token position during code generation based on immediate context and local complexity, which we term \ourapproach. The \ourapproach mechanism is demonstrated in Figure \ref{fig:intro_case}.

Realizing the \ourapproach mechanism presents significant challenges. Since LLMs do not spontaneously invoke reasoning during code generation, they must be explicitly taught this capability. We achieve this through cold-start training by constructing supervised learning samples that demonstrate reasoning invocation patterns of \ourapproach. While cold-start training can teach models to invoke reasoning blocks within code, it cannot effectively teach models where reasoning is necessary. The decision of which token positions to invoke thinking requires the model to identify its own moments of high complexity or logical risk, which demands adaptive judgment that goes beyond pattern matching in supervised data. To address this challenge, we employ Reinforcement Learning with Verifiable Rewards (RLVR) to enable LLMs to autonomously learn where to trigger reasoning during code generation, allowing the model to discover optimal thinking positions through reward-driven exploration.

In this work, we propose \ourapproach, a novel reasoning mechanism of LLMs for code generation that enables models to invoke thinking at any token position based on LLM's demands. \ourapproach is realized through a two-stage training pipeline. First, through cold-start training with carefully constructed code generation samples that demonstrate \ourapproach, we teach models the fundamental capability of pausing to think at arbitrary token positions during code generation. Second, we employ RLVR to further reinforce this capability, allowing models to autonomously explore and discover the optimal positions and strategies for invoking reasoning that suit the specific challenges they encounter. \ourapproach enables models to think on-demand at critical moments during code generation, precisely allocating computational resources to tokens that necessitate deep thinking. Moreover, by observing where and how models think during code generation, \ourapproach provides greater transparency into the decision-making process, enhancing the interpretability.

Extensive experiments demonstrate that \ourapproach achieves state-of-the-art performance over existing LLM reasoning-enhanced methods and recently proposed post-training methods on four mainstream code generation benchmarks, including LeetCode, LiveCodeBench, HumanEval, and MBPP. \ourapproach also exhibits strong generalization across different LLM families and model sizes. Ablation studies reveal that combining cold-start initialization with RLVR yields optimal results, and token-level thinking outperforms alternative variants such as line-level thinking. Further analysis highlights that LLMs tend to invoke thinking at positions with higher entropy, demonstrating that \ourapproach can reason at appropriate positions on demand.

\section{Related Work}

\paragraph{Reasoning and Planning Mechanisms in LLMs.}
Enhancing the reasoning and planning capabilities of LLMs has emerged as a central research focus in recent years.
A seminal advancement in this direction is Chain-of-Thought (CoT) prompting~\citep{wei2022chain}, which elicits complex reasoning by guiding LLMs to generate intermediate reasoning steps before arriving at a final answer.
Subsequent studies build on CoT with richer prompting strategies and search mechanisms~\citep{kojima2022zeroshot,wang2022selfconsistency,zhou2022leasttomost,yao2023treeofthoughts}.
In the domain of code generation, Self-Planning~\citep{jiang2024selfplanning} conducts problem decomposition and planning prior to code generation to reduce task complexity.
While these methods treat reasoning as an upfront thinking phase, recent work explores interleaved strategies that tightly couple thinking with task execution.
For instance, Interleaved Thinking~\citep{Xie2025InterleavedReasoningRL,liang2025plantain} guides LLMs to alternate between thinking and answering, enabling incremental refinement based on intermediate results. TwiG~\citep{guo2025twig} interleaves textual reasoning throughout visual generation trajectories, allowing reasoning to guide upcoming synthesis and reflect on previously generated content.

Recent advances in reasoning LLMs, such as DeepSeek-R1~\citep{guo2025deepseek} and Kimi-K2~\citep{Kimik2}, have achieved remarkable success by employing upfront thinking. While recent work on Interleaved Thinking allows reasoning to occur during implementation, it requires thinking at each sub-step and lacks the flexibility for on-demand invocation. This limitation introduces unnecessary computational overhead, while failing to allocate deeper reasoning effort to the most challenging portions of a task.

\paragraph{Post-Training of LLMs for Code Generation.}
Post-training has become important for improving the code generation capabilities of LLMs beyond pretraining, as it can better exploit task-specific data and verifiable execution signals.
One major approach is distillation from stronger reasoning LLMs. For example, OlympicCoder~\citep{openr1} fine-tunes models on competitive programming tasks using reasoning trajectories distilled from DeepSeek-R1. Similarly, OCR-Qwen-7B~\citep{ocrqwen} is distilled from DeepSeek-R1, leveraging a large-scale dataset of over 730K reasoning-annotated samples for open-source reproduction.
Another major approach is RL from executable feedback, which has been widely adopted to strengthen code generation and reasoning capabilities.
Skywork-OR1~\citep{skywork} employs large-scale RLVR training following DeepSeek-R1's pipeline for code generation.
CodePRM~\citep{codePRM} introduces a process reward model that provides step-level rewards for intermediate steps during generation.
CodeBoost~\citep{wang2025codeboost} enhances code generation through RL training on code reasoning tasks.
CodeRL+~\citep{jiang2025coderlplus} further enriches the learning signal by aligning code generation with execution semantics beyond binary pass/fail feedback.

Existing post-training methods, regardless of whether they are based on distillation or RL, predominantly adopt the upfront thinking practice. This introduces the limitations discussed in Section~\ref{sec:intro}, necessitating a shift in the thinking approach for code generation.

\section{Methodology}

\subsection{Defining \ourapproach}

We begin by formally defining the \ourapproach mechanism and contrasting it with the conventional upfront thinking method.
Let $x$ denote the requirement and $c$ denote the generated code.  We define two special token pairs: $\langle \texttt{think} \rangle$ and $\langle \texttt{/think} \rangle$ for the upfront thinking block, and $\langle \texttt{thinkanywhere} \rangle$ and $\langle \texttt{/thinkanywhere} \rangle$ for the \ourapproach thinking block.

\paragraph{Upfront Thinking.}
In the upfront thinking method adopted by existing reasoning-enhanced LLMs~\citep{jaech2024o1systemcard,guo2025deepseek}, the generation process can be decomposed into two sequential phases. Given input $x$, the model first generates a complete reasoning trace $s$ enclosed within $\langle \texttt{think} \rangle$ and $\langle \texttt{/think} \rangle$ tokens, and then generates the code $c$ conditioned on both $x$ and $s$:
\begin{equation}
    P(c, s \mid x) = \underbrace{P(s \mid x)}_{\text{upfront reasoning}} \cdot \underbrace{P(c \mid x, s)}_{\text{code generation}}.
\end{equation}
This formulation enforces a strict separation between reasoning and code generation, making LLM difficult to invoke additional reasoning in code generation process.

\paragraph{\ourapproachbf.}
\ourapproach enables LLM to precisely reason at any position where deliberation is needed during code generation. Considering the non-uniform distribution of logical complexity in code generation, \ourapproach allows the model to dynamically scale its reasoning length at challenging bottlenecks, achieving a truly on-demand allocation of computational resources.
Formally, the model generates a mixed sequence $\mathbf{y}$. This sequence naturally decomposes into code segments and thinking blocks:
\begin{equation}
    \mathbf{y} = (s, c^{(1)}, h^{(1)}, c^{(2)}, h^{(2)}, \ldots, c^{(M)}, h^{(M)}, c^{(M+1)}),
\end{equation}
where $s$ denotes the initial thinking block enclosed within $\langle \texttt{think} \rangle$ and $\langle \texttt{/think} \rangle$, each $c^{(i)}$ represents a code segment, and each $h^{(i)}$ represents a thinking block enclosed within $\langle \texttt{thinkanywhere} \rangle$ and $\langle \texttt{/thinkanywhere} \rangle$ tokens that is placed between code segments. The number of thinking blocks $M \geq 0$ and their positions are dynamically determined by the model during generation.

The generation process of \ourapproach can be formulated as:
\begin{equation}
\begin{split}
 P (\mathbf{y} \mid x) = P(s \mid x) \cdot \prod_{i=1}^{M} \Big[ P(c^{(i)} \mid x, \mathbf{y}_{<c^{(i)}}) \cdot P(h^{(i)} \mid x, \mathbf{y}_{<h^{(i)}}) \Big] \cdot P(c^{(M+1)} \mid x, \mathbf{y}_{<c^{(M+1)}}),
\end{split}
\end{equation}
where $\mathbf{y}_{<c^{(i)}}$ and $\mathbf{y}_{<h^{(i)}}$ denote all preceding tokens before code segment $c^{(i)}$ and thinking block $h^{(i)}$, respectively. Notably, upfront thinking can be viewed as a special case of \ourapproach where thinking occurs exclusively at the beginning.

The final executable code $c$ is obtained by removing all thinking blocks from $\mathbf{y}$, including the initial $\langle \texttt{think} \rangle$ block and all $\langle \texttt{thinkanywhere} \rangle$ blocks:
\begin{equation}
    c = c^{(1)} \oplus c^{(2)} \oplus \cdots \oplus c^{(M+1)},
\end{equation}
where $\oplus$ denotes sequence concatenation.

\begin{table*}[t]
\centering
\small
\begin{tabular}{p{0.95\linewidth}}
\toprule
You are a coding assistant that generates both code and inline self-guidance signals. First output \texttt{<think>...
</think>} with brief reasoning, then output the final code. \\
\\
MUST FOLLOW Rules for \texttt{<thinkanywhere>...</thinkanywhere>} tags: \\
1. You MUST use \texttt{<thinkanywhere>...</thinkanywhere>} tags for self-guidance or intermediate reasoning. \\
2. \texttt{<thinkanywhere>...</thinkanywhere>} MUST be embedded within an existing program statement token sequence. \\
3. The code must remain valid and executable after removing all \texttt{<thinkanywhere>...</thinkanywhere>} segments. \\
\\
User: \textcolor{red}{Prompt}. Assistant: \\
\bottomrule
\end{tabular}
\caption{Template for \ourapproach. \textcolor{red}{Prompt} will be replaced with the specific coding requirement.}
\label{tab:template}
\end{table*}

\paragraph{Training Template.}
To train \ourapproach, we design a template that guides LLMs to follow the \ourapproach generation format, as shown in Table~\ref{tab:template}. The template instructs the model to first produce initial reasoning within $\langle \texttt{think} \rangle$ tags, then generate code with $\langle \texttt{thinkanywhere} \rangle$ blocks invoked at positions requiring deliberation. We constrain only the structural format while avoiding content-specific biases, allowing the model to discover optimal thinking patterns through subsequent reinforcement learning.

\subsection{Cold Start for \ourapproach}

LLMs do not invoke thinking blocks during code generation, and even explicit instructions in prompts often fail to enforce this behavior reliably. Therefore, they must be explicitly taught this capability through training. The goal of cold start is to equip the model with the fundamental ability to reason at arbitrary positions within code.

\paragraph{Automatic Data Construction.}
We leverage strong reasoning LLMs with our training template to automatically construct training data that demonstrates the \ourapproach pattern. Specifically, we prompt the reasoning LLMs to solve coding problems while explicitly invoking thinking blocks enclosed within $\langle \texttt{thinkanywhere} \rangle$ and $\langle \texttt{/thinkanywhere} \rangle$ tokens at positions where deliberation is needed during code generation.

To ensure data quality, we filter out samples with incorrect formatting, such as malformed thinking block boundaries or improper nesting of special tokens. Following prior work~\citep{codeio} that demonstrates both correct and incorrect solutions contribute to model learning, we retain samples regardless of code correctness. This process of data construction yields approximately 5,000 training samples.

We perform supervised fine-tuning using LoRA~\citep{lora} on the constructed training samples as cold start. Following~\citep{schulman2025lora}, we adopt LoRA over full-parameter SFT as it achieves comparable performance with greater robustness and lower computational overhead. This training enables the model to learn the pattern of invoking $\langle \texttt{thinkanywhere} \rangle$ blocks within code, acquiring the basic capability that serves as the foundation for subsequent reinforcement learning.

\paragraph{Dedicated Reasoning Trigger Token.}
In default implementation, $\langle \texttt{thinkanywhere} \rangle$ is tokenized into multiple ordinary tokens, each carrying its own lexical meaning. Requiring the model to use these tokens simultaneously as lexical units and as a trigger signal for invoking reasoning introduces semantic ambiguity. Moreover, generating a multi-token delimiter increases the prediction path length for a single control decision, making the trigger less reliable. We therefore introduce a special token variant (\ourapproachstar) that represents the thinking delimiter as a single dedicated vocabulary entry, providing an unambiguous and efficient signal for invoking inline reasoning.

However, directly adding randomly initialized special tokens is ineffective, as the limited post-training data is insufficient for the model to learn meaningful representations from scratch. To address this, we propose a semantic-aware initialization strategy that composes the embedding from two complementary sources: the semantic content of the trigger and the structural role of a delimiter. Specifically, we initialize the embeddings of the new special tokens as:
\begin{align}
    \mathbf{e}_{\langle \texttt{ta} \rangle} &= 0.5 \cdot \text{mean}(\mathbf{e}_\texttt{think}, \mathbf{e}_\texttt{any}, \mathbf{e}_\texttt{where}) + 0.5 \cdot \mathbf{e}_{\langle \texttt{im\_start} \rangle}, \\
    \mathbf{e}_{\langle \texttt{/ta} \rangle} &= 0.5 \cdot \text{mean}(\mathbf{e}_\texttt{think}, \mathbf{e}_\texttt{any}, \mathbf{e}_\texttt{where}) + 0.5 \cdot \mathbf{e}_{\langle \texttt{im\_end} \rangle},
\end{align}
where $\mathbf{e}_{\langle \texttt{ta} \rangle}$ and $\mathbf{e}_{\langle \texttt{/ta} \rangle}$ denote the embeddings of the opening and closing special tokens, respectively. The first term encodes the semantic intent of ``think anywhere,'' while the second term inherits the structural behavior of existing delimiter tokens ($\langle \texttt{im\_start} \rangle$ and $\langle \texttt{im\_end} \rangle$), which the model has already learned to treat as mode-switching boundaries during pretraining.

To effectively train the dedicated trigger tokens, we adopt a two-stage cold-start procedure:
\begin{enumerate}
    \item \textbf{Stage 1: Embedding alignment.} We freeze the model parameters and train only the input embeddings and LM head weights. This stage allows the tokens to develop appropriate representations without disrupting the model's existing capabilities.
    \item \textbf{Stage 2: Joint fine-tuning.} We continue training the special token embeddings and LM head jointly with LoRA adapters applied to the model, enabling the model to learn how to generate and respond to the dedicated trigger tokens in context.
\end{enumerate}
The subsequent RLVR stage proceeds identically to the default version.

\subsection{RLVR for \ourapproach}

We then employ RLVR to enable the LLMs to autonomously discover optimal thinking positions and strategies through reward-driven exploration.

\paragraph{Reinforcement Learning Algorithm.}
We adopt Group Relative Policy Optimization (GRPO)~\citep{DeepSeekMath} as our reinforcement learning algorithm. Unlike Proximal Policy Optimization (PPO)~\citep{PPO} which requires a separate value model to estimate baselines, GRPO computes baselines from group-level statistics, eliminating the need for an additional value model and significantly reducing computational overhead.

Specifically, for each input $x$, GRPO samples a group of $G$ candidate outputs $\{y_1, y_2, \ldots, y_G\}$ from the current policy $\pi_\theta$. The reward for each output $y_i$ is computed as $R(y_i)$, and the group-normalized advantage is calculated as:
\begin{equation}
    \hat{A}_i = \frac{R(y_i) - \text{mean}(\{R(y_j)\}_{j=1}^{G})}{\text{std}(\{R(y_j)\}_{j=1}^{G})}.
\end{equation}
The policy is then optimized by maximizing the clipped surrogate objective with a KL divergence penalty:
\begin{align}
    \mathcal{L}_{\text{GRPO}}(\theta) = \mathbb{E} \Biggl[ \min \left( \rho_i \hat{A}_i, \text{clip}(\rho_i, 1-\epsilon, 1+\epsilon) \hat{A}_i \right) - \beta \cdot D_{\text{KL}}(\pi_\theta \| \pi_{\text{ref}}) \Biggr],
\end{align}
where $\rho_i = \frac{\pi_\theta(y_i \mid x)}{\pi_{\text{old}}(y_i \mid x)}$ denotes the probability ratio, $\epsilon$ is the clipping threshold, and $\beta$ controls the strength of the KL penalty against the reference policy $\pi_{\text{ref}}$.

\paragraph{Reward Modeling.}
We design a hierarchical reward function for \ourapproach. The reward consists of two components: a reasoning structure reward $R_{\text{struct}}$ and a code correctness reward $R_{\text{correct}}$, combined in a gated manner:
\begin{align}
    R(y) = \alpha \cdot R_{\text{struct}}(y) + (1-\alpha) \cdot R_{\text{correct}}(y),
\end{align}
where $\alpha = 0.1$ controls the weight between the two components.

The reasoning structure reward $R_{\text{struct}} \in \{0, 1\}$ verifies that the model adheres to the \ourapproach reasoning definition. Specifically, it checks whether the output contains an initial thinking block within $\langle \texttt{think} \rangle$ and $\langle \texttt{/think} \rangle$ tags, followed by code that incorporates $\langle \texttt{thinkanywhere} \rangle$ blocks:
\begin{align}
    R_{\text{struct}}(y) = \mathbbm{1} \left[ \texttt{HasInitialThinking}(y) \land \texttt{HasThinkAnywhere}(y) \right],
\end{align}
where $\texttt{HasInitialThinking}(\cdot)$ verifies the presence of the initial $\langle \texttt{think} \rangle$ block, and $\texttt{HasThinkAnywhere}(\cdot)$ ensures that at least one $\langle \texttt{thinkanywhere} \rangle$ block is embedded within the generated code. This reward encourages the model to actively engage in on-demand reasoning throughout the generation process.

The code correctness reward $R_{\text{correct}} \in \{0, 1\}$ evaluates the functional correctness of the generated code by executing it against the provided test cases:
\begin{equation}
    R_{\text{correct}}(y) = \mathbbm{1} \left[ \texttt{PassAllTests}(c) \right].
\end{equation}


\begin{table*}[t]
\centering
\caption{Performance of \ourapproach compared to post-training methods and reasoning-enhanced methods. Best results are in \textbf{bold}. \ourapproachstar denotes the special token variant with semantic-aware initialization (see Section~3.2).}
\label{tab:main_results}
\small
{
\begin{tabular}{lccccc}
\toprule
\textbf{Method} & \textbf{LeetCode} & \textbf{LiveCodeBench} & \textbf{HumanEval} & \textbf{MBPP} & \textbf{Average} \\
\midrule
Base Model & 50.6 & 34.3 & 88.4 & 70.7 & 61.0 \\
\midrule
\multicolumn{6}{l}{\textbf{Post-Training Methods}} \\
OlympicCoder & 45.3 & 30.9 & 75.6 & 67.2 & 54.8 \\
OCR-Qwen-7B & 53.3 & 33.0 & 86.8 & 58.9 & 58.0 \\
CodePRM & 52.8 & 34.8 & 88.4 & 73.9 & 62.5 \\
CodeBoost & 53.3 & 34.6 & 87.2 & 65.7 & 60.2 \\
CodeRL+ & 63.3 & 36.9 & 90.9 & 76.2 & 66.8 \\
\midrule
\multicolumn{6}{l}{\textbf{Reasoning-Enhanced Methods}} \\
CoT & 53.9 & 30.9  & 86.6 & 77.7  & 62.3 \\
Self-planning  & 49.2 & 31.1  & 86.9 & 77.9 & 61.3 \\
Interleaved Thinking & 50.6 & 30.7  & 86.4 & 79.2 & 61.7 \\
GRPO & 67.3 & 36.0 & 88.6 & 81.7 & 68.4 \\
\midrule
\ourapproach (Prompting)  & 41.1 & 34.4 & 84.8 & 67.4 & 56.9 \\
\ourapproachstar (SFT) & 46.7 & 32.5 & 79.9 & 78.2 & 59.3 \\
\ourapproachstar (Ours) & 68.9 & 36.7 & 90.2 & 84.5 & 70.0 \\
\ourapproach (SFT) & 47.9 & 32.3 & 82.9 & 79.4 & 60.6 \\
\textbf{\ourapproach (Ours)} & \textbf{69.4} & \textbf{37.2} & \textbf{91.5} & \textbf{82.9} & \textbf{70.3} \\
\bottomrule
\end{tabular}
}
\end{table*}


\section{Experiments}

\subsection{Experiment Setup}

\paragraph{Training Details.}
Follow previous work~\citep{skywork}, our training corpus comprises 14K programming problem from the Skywork dataset. By default, we employ Qwen2.5-Coder-7B-Instruct~\citep{hui2024qwen2} as the base model for our experiments. The RL algorithm is implemented using the VeRL framework~\citep{sheng2024hybridflow}. Training parameters are set as follows: batch size of 128, mini-batch size of 64, learning rate of $1 \times 10^{-6}$, and 2 training epochs. Each problem generates 8 rollout samples up to 4096 tokens. The experiments run on 8 NVIDIA A100 GPUs (40G). We employ Google's Gemini 2.5 Flash~\citep{gemini2.5} to synthesize cold-start training data.

\paragraph{Evaluation Details.}
Following established practices in prior work~\citep{codePRM,codereasoner,wang2025codeboost,DBLP:journals/corr/abs-2508-00222,DBLP:conf/acl/DongJLJGYL24}, our evaluation encompasses four widely-used code generation benchmarks: HumanEval~\citep{chen2021evaluating}, MBPP~\citep{mbpp}, LeetCode~\citep{leetcode}, and LiveCodeBench~\citep{jain2024livecodebench}. We adopt pass@1 as our primary evaluation metric. To ensure reproducibility and consistency across all experiments, we employ greedy sampling with the temperature fixed at 0.

\paragraph{Baselines.}
Beyond the base model and standard GRPO method~\citep{DeepSeekMath}, we compare \ourapproach with two categories of methods, all using the same base model.
The first category includes the reasoning-enhanced approaches that incorporate thinking mechanisms, including \textbf{CoT}~\citep{wei2022chain}, \textbf{Self-Planning}~\citep{jiang2024selfplanning}, and \textbf{Interleaved Thinking}~\citep{Xie2025InterleavedReasoningRL}\footnote{As Interleaved Thinking does not provide source code, we adapt it to the code generation setting by prompting the model to alternate between reasoning and code implementation, following the method described in the original work.}.
The second category includes the recently proposed post-training models and methods developed for code generation, including \textbf{OlympicCoder}~\citep{openr1}, \textbf{OCR-Qwen-7B}~\citep{ocrqwen}, \textbf{CodePRM}~\citep{codePRM}, \textbf{CodeBoost}~\citep{wang2025codeboost}, and \textbf{CodeRL+}~\citep{jiang2025coderlplus}.

\subsection{Experiment Results}

\paragraph{Performance of \ourapproachbf.}
Table~\ref{tab:main_results} presents the main results of \ourapproach compared to baselines on four benchmarks. Overall, \ourapproach achieves the best performance across all benchmarks, with an average score of 70.3\%, representing a 9.3\% absolute improvement over the base model.
Compared to Post-Training Methods, \ourapproach surpasses the best-performing baseline CodeRL+, demonstrating the effectiveness of our approach over other RL-based approaches.
Compared to Reasoning-Enhanced Methods, \ourapproach substantially outperforms CoT, Self-planning, Interleaved Thinking, and GRPO across all metrics. Notably, some methods exhibit inconsistent improvements across different benchmarks. In contrast, \ourapproach achieves consistent improvements on both simple and challenging tasks, suggesting that our dynamic thinking strategy is more effective than fixed reasoning patterns.
Furthermore, the comparison among \ourapproach variants reveals the importance of RL-based training. \ourapproach (Prompting) and \ourapproach (SFT) underperform the base model on several benchmarks, whereas \ourapproach with RL training achieves substantial gains, highlighting that reinforcement learning is crucial for learning effective thinking patterns.

We also report the results of the special token variant (\ourapproachstar). With semantic-aware initialization and the two-stage cold-start procedure, \ourapproachstar achieves an average score of 70.0\%, which is comparable to the default text-based version (70.3\%). 
We observe that the text-based version tends to invoke thinking blocks at stereotyped positions (\eg, after ``='' tokens), while the special token variant exhibits more diverse and contextually appropriate placement. However, the limited post-training data constrains the special token variant from fully learning the semantics of the new tokens. We believe that natively integrating \ourapproach special tokens during large-scale pretraining would further unlock their potential. Since the text-based version slightly outperforms the special token variant under our post-training setting, we adopt the text-based version for all subsequent experiments.

\paragraph{Cross-Domain Generalization.}
To investigate whether \ourapproach generalizes beyond code generation, we directly evaluate our code-domain-trained model on mathematical reasoning benchmarks, including AIME 2024~\citep{aime24}, AIME 2025~\citep{aime25}, and HMMT 2025~\citep{hmmt25}. Table~\ref{tab:math} reports the results under pass@1, pass@5, and pass@10 settings. Notably, although \ourapproach is trained exclusively on code generation tasks, it achieves consistent and substantial improvements over both the base model and GRPO across three mathematical reasoning benchmarks. For instance, on AIME 2024, \ourapproach improves pass@1 from 5.3\% (Base Model) and 6.0\% (GRPO) to 17.3\%, representing a remarkable gain. Similar trends are observed on AIME 2025 and HMMT 2025, where \ourapproach achieves 17.7\% and 14.4\% pass@1 respectively. These results suggest that the think-on-demand reasoning capability acquired through \ourapproach is not domain-specific but transfers across tasks, demonstrating strong cross-domain generalization.

\begin{table}[t]
\centering
\caption{Cross-domain generalization of \ourapproach to mathematical reasoning benchmarks. The model is trained solely on code generation tasks.}
\label{tab:math}
\small
\resizebox{\columnwidth}{!}{
\begin{tabular}{lccccccccc}
\toprule
\multirow{2}{*}{\textbf{Method}} & \multicolumn{3}{c}{\textbf{AIME 2024}} & \multicolumn{3}{c}{\textbf{AIME 2025}} & \multicolumn{3}{c}{\textbf{HMMT 2025}} \\
\cmidrule(lr){2-4} \cmidrule(lr){5-7} \cmidrule(lr){8-10}
& pass@1 & pass@5 & pass@10 & pass@1 & pass@5 & pass@10 & pass@1 & pass@5 & pass@10 \\
\midrule
Base Model & 5.3 & 14.6 & 20.0 & 4.0 & 13.4 & 16.7 & 0.0 & 0.0 & 0.0 \\
GRPO & 6.0 & 16.8 & 23.3 & 4.7 & 17.2 & 26.7 & 0.3 & 1.7 & 3.3 \\
\textbf{\ourapproach} & \textbf{17.3} & \textbf{32.9} & \textbf{40.2} & \textbf{17.7} & \textbf{28.0} & \textbf{33.2} & \textbf{14.4} & \textbf{18.5} & \textbf{19.6} \\
\bottomrule
\end{tabular}
}
\end{table}

\begin{wraptable}{r}{0.49\textwidth}
\centering
\vspace{-12pt}
\caption{Generalizability of \ourapproach across different model families and scales.}
\label{tab:generalization}
\small{
\begin{tabular}{lcc}
\toprule
\textbf{Model} & \textbf{Average} & \textbf{$\Delta$ vs.\ Base} \\
\midrule
\multicolumn{3}{l}{\textbf{Qwen2.5-Coder-7B-Instruct}} \\
~~Base Model & 61.0 & -- \\
~~+ GRPO & 68.4 & +7.4 \\
~~+ \ourapproach & \textbf{70.3} & \textbf{+9.3} \\
\midrule
\multicolumn{3}{l}{\textbf{Qwen2.5-Coder-1.5B-Instruct}} \\
~~Base Model & 40.6 & -- \\
~~+ GRPO & 51.9 & +11.3 \\
~~+ \ourapproach & \textbf{54.5} & \textbf{+13.9} \\
\midrule
\multicolumn{3}{l}{\textbf{LLaMA-3.1-8B-Instruct}} \\
~~Base Model & 38.4 & -- \\
~~+ GRPO & 42.0 & +3.6 \\
~~+ \ourapproach & \textbf{43.8} & \textbf{+5.4} \\
\bottomrule
\end{tabular}
}
\vspace{-10pt}
\end{wraptable}

\paragraph{Application on Various LLMs.}
To validate the generalizability of \ourapproach, we evaluated its performance on three diverse LLMs spanning different model families and parameter scales: LLaMA-3.1-8B-Instruct~\citep{llama31}, Qwen2.5-Coder-7B-Instruct~\citep{hui2024qwen2}, and Qwen2.5-Coder-1.5B-Instruct~\citep{hui2024qwen2}. Table~\ref{tab:generalization} reports the average performance across four benchmarks. The results demonstrate that \ourapproach consistently outperforms both the base model and GRPO across all LLMs, with substantial margins over GRPO. Notably, \ourapproach achieves up to +13.9\% improvement over the base model. Furthermore, \ourapproach exhibits strong scalability across different model sizes. On the smaller Qwen2.5-Coder-1.5B-Instruct, \ourapproach achieves a substantial improvement over the base model, indicating that our method is particularly effective for smaller models with limited capacity.

\subsection{Ablation Study}

To understand the contribution of each component in \ourapproach, we conduct comprehensive ablation studies comparing multiple variants on the LeetCode benchmark. We first ablate different training strategies: 1) \textbf{\ourapproach}: Our complete method incorporates both cold-start training and RLVR in a two-stage pipeline. 2) \textbf{Only Cold Start}: Model trained solely with supervised learning on annotated samples of \ourapproach, without RL phase. 3) \textbf{Only RLVR}: Model trained directly with RLVR of \ourapproach from scratch, bypassing the cold-start phase. 4) \textbf{Line-level Thinking}: A variant where RLVR encourages line-level thinking (similar to comment-style reasoning) rather than arbitrary token positions. 5) \textbf{No Upfront Thinking}: A variant of our approach that removes the initial thinking block and relies solely on \ourapproach within the code. To isolate the impact of the \ourapproach mechanism itself, we evaluate an inference variant: 6) \textbf{Padding Thinking}: During \ourapproach generation, the content within \texttt{<thinkanywhere>} blocks is replaced with padding tokens before continuing generation.

\begin{wraptable}{r}{0.42\textwidth}
\centering
\vspace{-12pt}
\caption{The results of ablation study.}
\label{tab:ablation}
\small{
\begin{tabular}{lcc}
\toprule
\textbf{Method} & \textbf{Pass@1} & \textbf{$\Delta$} \\
\midrule
\textbf{\ourapproach} & \textbf{69.4} & \textbf{--} \\
\midrule
~~Only Cold Start & 47.9 & -21.5 \\
~~Only RLVR & 63.4 & -6.0 \\
\hdashline
~~Line-level Thinking & 67.2 & -2.2 \\
~~No Upfront Thinking & 66.6 & -2.8 \\
~~Padding Thinking & 67.6 & -1.8 \\
\bottomrule
\end{tabular}}
\vspace{-10pt}
\end{wraptable}

The results are presented in Table~\ref{tab:ablation}. We have the following observations. First, both cold-start and RLVR are essential. Removing either training stage leads to substantial performance degradation. Only Cold Start performs poorly, indicating that supervised learning alone is insufficient for the model to learn effective thinking. Only RLVR performs better but still lags behind the full method, suggesting that cold-start initialization helps stabilize RL training. Second, Line-level Thinking underperforms our token-level approach, suggesting that restricting thinking to line boundaries limits the model's ability to invoke reasoning at optimal positions, validating our design choice of allowing thinking at arbitrary token positions.
Third, No Upfront Thinking incurs only a moderate drop (-2.8\%), indicating that the primary performance gains of \ourapproach stem from the \ourapproach mechanism within the code rather than the upfront thinking phase. Finally, Padding Thinking also shows moderate performance degradation, demonstrating that the reasoning content within \texttt{<thinkanywhere>} blocks is indeed valuable. However, the performance does not fully deteriorate to the base model level, suggesting that identifying appropriate thinking positions is also important. Through the subsequent padding tokens, the model still performs some implicit reasoning during the forward pass~\citep{goyal2024think,pfau2024lets}.

\subsection{Further Analysis}

\begin{figure}[t]
    \centering
    \begin{subfigure}{0.4\linewidth}
        \centering
        \includegraphics[width=\linewidth]{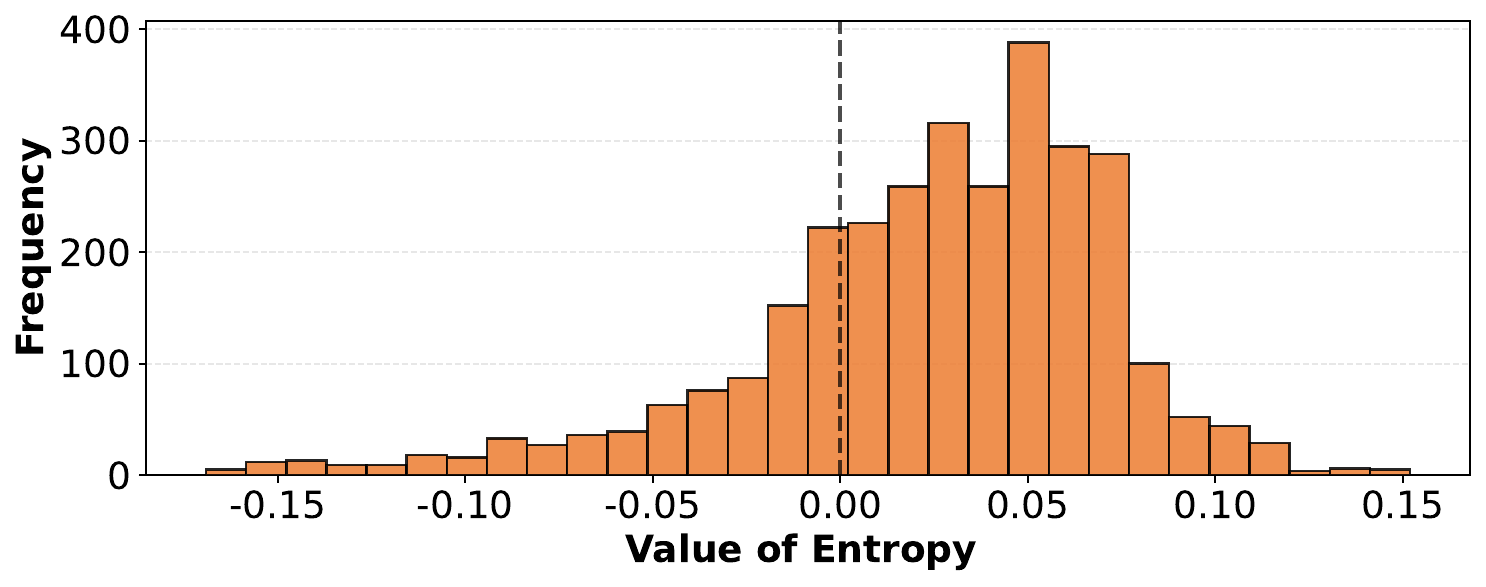}
        \caption{Token entropy analysis.}
        \label{fig:entropy_distribution}
    \end{subfigure}
    \qquad
    \begin{subfigure}{0.4\linewidth}
        \centering
        \includegraphics[width=\linewidth]{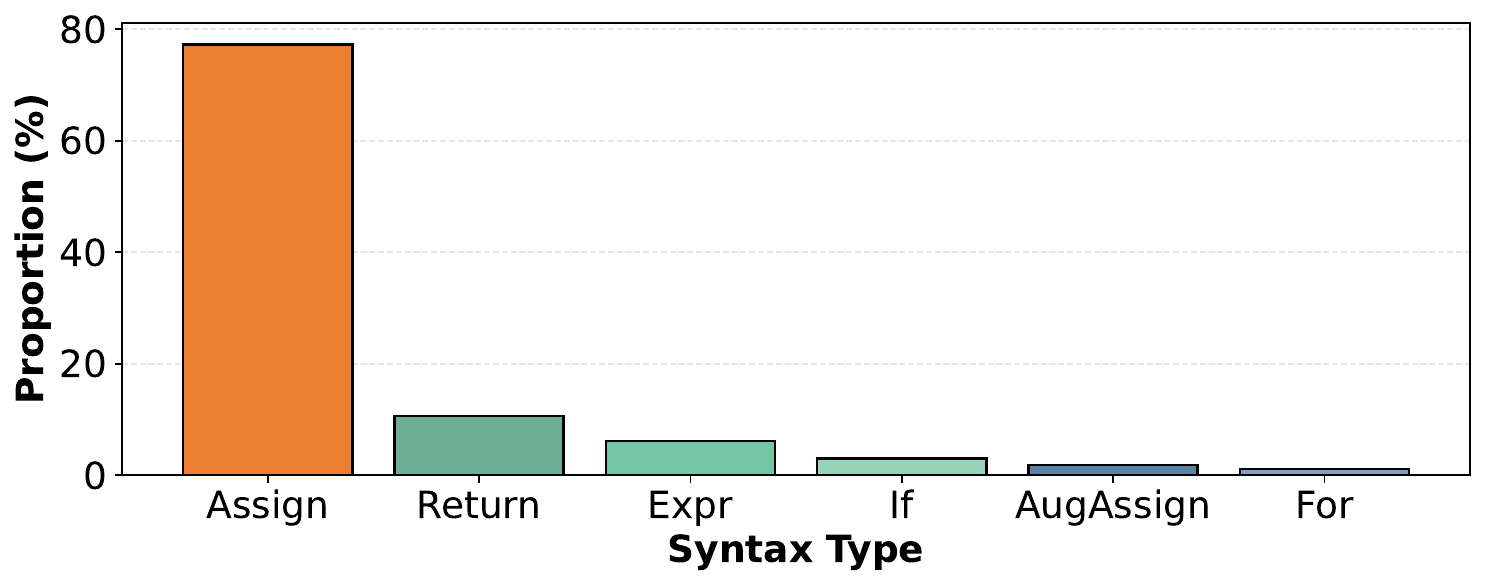}
        \caption{Syntactic context analysis.}
        \label{fig:syntax_distribution_final}
    \end{subfigure}
    \caption{Results of Thinking Position Analysis.}
    \label{fig:combined}
\end{figure}

\paragraph{Thinking Position Analysis.}
Understanding where \ourapproach chooses to invoke reasoning during code generation provides crucial insights into the model's perception of code complexity and validates whether it truly allocates computational resources efficiently. We analyze generated solutions on the LeetCode benchmark through two perspectives: 1) Token entropy analysis: We compute the average token entropy over the n tokens following each \texttt{<thinkanywhere>} block and compare it against a baseline where no thinking blocks are generated, thereby quantifying the impact of \texttt{<thinkanywhere>} on entropy. We empirically set n=10 for entropy analysis, as this window size typically captures a statement unit, thereby mitigating individual token noise. 2) Syntactic context analysis: We employ an AST parser to identify the syntactic category of the statement enclosing each thinking position (e.g., If, While, FunctionDef, BinOp), characterizing where the model chooses to think within the code structure.

Figure~\ref{fig:entropy_distribution} shows the distribution of entropy differences between thinking-disabled/enabled runs at positions where the model originally invoked <thinkanywhere>. We observe that the differences are predominantly positive, indicating higher entropy when thinking is disabled. This suggests that the model tends to invoke <thinkanywhere> at positions where it anticipates high uncertainty, effectively identifying challenging points in code generation.
Figure~\ref{fig:syntax_distribution_final} presents the top five syntactic categories where <thinkanywhere> is invoked. The model most frequently invokes thinking at assignment statements, likely because assignments often involve complex computations or variable updates that benefit from intermediate reasoning. Return statements rank second, which we attribute to the model's tendency to deliberate on final outputs to ensure correctness before concluding a function.

\paragraph{Computational Efficiency Comparison.}

\begin{wrapfigure}{r}{0.5\textwidth}
    \centering
    \includegraphics[width=0.5\textwidth]{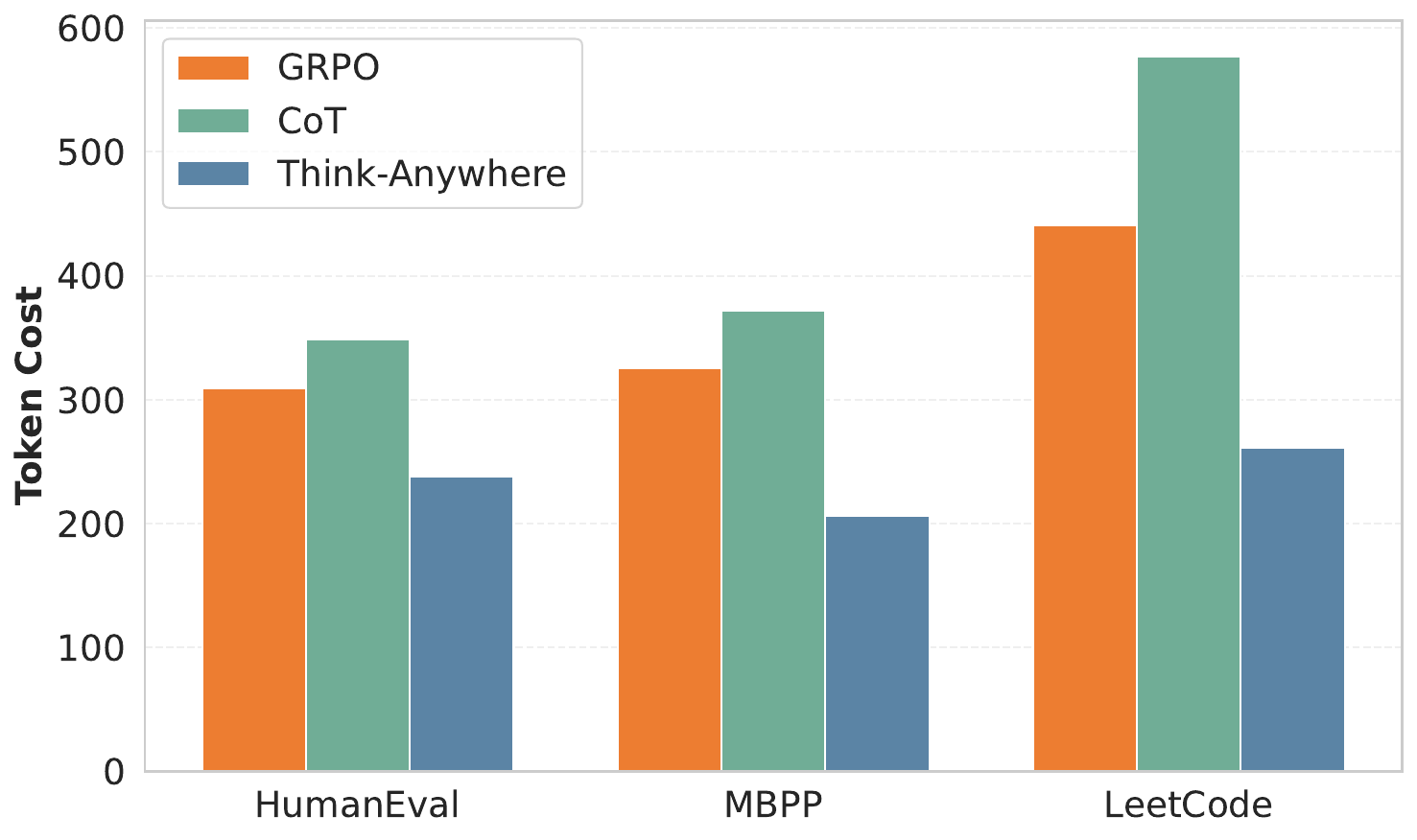}
    \caption{Token cost of different methods.}
    \label{fig:token_cost}
    \vspace{-10pt}
\end{wrapfigure}

We evaluate the inference efficiency of \ourapproach by measuring the average number of tokens generated per solution. We compare \ourapproach against two reasoning baselines: GRPO (upfront thinking) and CoT prompting. As shown in Figure~\ref{fig:token_cost}, \ourapproach consistently generates fewer tokens than both baselines across benchmarks. The reduction in total token cost is attributed to the fact that \ourapproach shortens the upfront thinking phase while introducing additional \texttt{<thinkanywhere>} tokens on demand. Since GRPO and CoT can only reason before code generation, they are forced to deliberate exhaustively at the upfront thinking stage, anticipating all potential implementation challenges upfront, which results in lengthy reasoning traces. \ourapproach, by contrast, invokes deliberation where it is needed. The upfront thinking phase therefore only needs to handle high-level planning, and its length is substantially reduced. The token savings from the shortened upfront thinking far outweigh the cost of the additional \texttt{<thinkanywhere>} blocks, resulting in a net reduction in total token usage. A detailed breakdown of the upfront thinking length and \texttt{<thinkanywhere>} block length is provided in Appendix~\ref{appendix:token_breakdown}.

\begin{figure}[h]
    \centering
    \includegraphics[width=\textwidth]{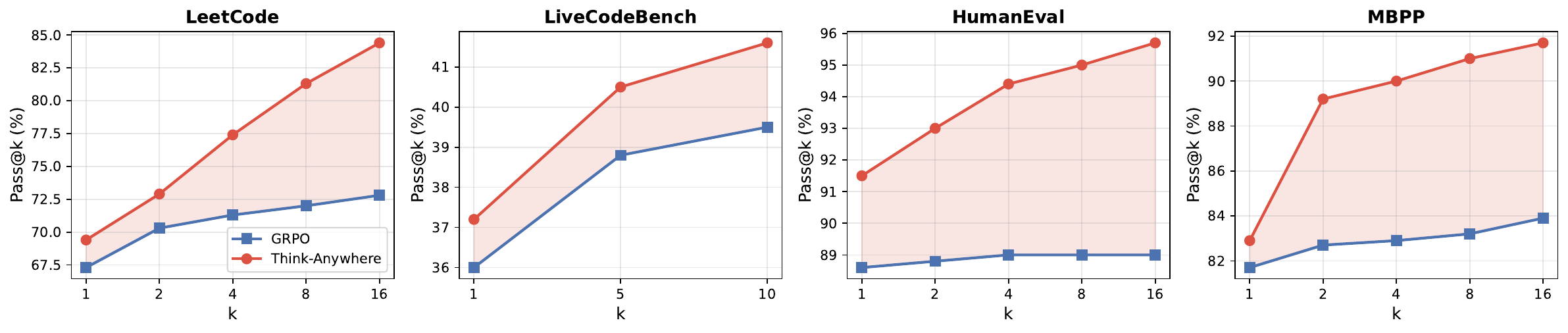}
    \caption{Pass@k comparison between GRPO and \ourapproach across four benchmarks.}
    \label{fig:passk}
\end{figure}

\paragraph{Pass@k Analysis.}

Pass@k reflects the upper bound of a model's capability by evaluating whether at least one correct solution exists among $k$ sampled candidates. We report pass@k results for both GRPO and \ourapproach across all benchmarks to investigate whether \ourapproach expands the model's capability boundary. As shown in Figure~\ref{fig:passk}, \ourapproach consistently outperforms GRPO across all values of $k$ on benchmarks. More importantly, the performance gap between \ourapproach and GRPO widens significantly as $k$ increases, particularly on LeetCode and MBPP. This widening gap demonstrates that \ourapproach substantially raises the model's capability ceiling.

\section{Conclusion}

In this work, we introduce \ourapproach, a novel reasoning mechanism that enables LLMs to invoke thinking at any token position during code generation. Unlike conventional upfront thinking approaches that enforce a strict separation between reasoning and code implementation, \ourapproach allows models to deliberate precisely where complexity arises. Extensive experiments across multiple mainstream benchmarks demonstrate that \ourapproach achieves SOTA performance, with strong generalization across different LLMs. Beyond performance gains, our analysis reveals that LLMs naturally learn to invoke thinking at high-entropy positions, suggesting that \ourapproach enables adaptive computation where reasoning effort is dynamically allocated based on local complexity.
We believe \ourapproach opens promising directions for future research, including extending \ourapproach to other domains beyond code generation, and investigating how models can learn what \emph{not} to think, further optimizing the trade-off between reasoning depth and computational efficiency.

\bibliography{custom}

@article{schulman2025lora,
  author = {John Schulman and Thinking Machines Lab},
  title = {LoRA Without Regret},
  journal = {Thinking Machines Lab: Connectionism},
  year = {2025},
  note = {https://thinkingmachines.ai/blog/lora/},
  doi = {10.64434/tml.20250929},
}

@article{jiang2024selfplanning,
  author       = {Xue Jiang and
                  Yihong Dong and
                  Lecheng Wang and
                  Zheng Fang and
                  Qiwei Shang and
                  Ge Li and
                  Zhi Jin and
                  Wenpin Jiao},
  title        = {Self-Planning Code Generation with Large Language Models},
  journal      = {{ACM} Trans. Softw. Eng. Methodol.},
  volume       = {33},
  number       = {7},
  pages        = {182:1--182:30},
  year         = {2024}
}

@misc{guo2025twig,
      title={Thinking-while-Generating: Interleaving Textual Reasoning throughout Visual Generation}, 
      author={Ziyu Guo and Renrui Zhang and Hongyu Li and Manyuan Zhang and Xinyan Chen and Sifan Wang and Yan Feng and Peng Pei and Pheng-Ann Heng},
      year={2025},
      eprint={2511.16671},
      archivePrefix={arXiv},
      primaryClass={cs.CV},
      url={https://arxiv.org/abs/2511.16671}, 
}

@article{jaech2024o1systemcard,
  author       = {Aaron Jaech and
                  Adam Kalai and
                  Adam Lerer and
                  Adam Richardson and
                  Ahmed El{-}Kishky and
                  Aiden Low and
                  Alec Helyar and
                  Aleksander Madry and
                  Alex Beutel and
                  Alex Carney and
                  Alex Iftimie and
                  Alex Karpenko and
                  Alex Tachard Passos and
                  Alexander Neitz and
                  Alexander Prokofiev and
                  Alexander Wei and
                  Allison Tam and
                  Ally Bennett and
                  Ananya Kumar and
                  Andre Saraiva and
                  Andrea Vallone and
                  Andrew Duberstein and
                  Andrew Kondrich and
                  Andrey Mishchenko and
                  Andy Applebaum and
                  Angela Jiang and
                  Ashvin Nair and
                  Barret Zoph and
                  Behrooz Ghorbani and
                  Ben Rossen and
                  Benjamin Sokolowsky and
                  Boaz Barak and
                  Bob McGrew and
                  Borys Minaiev and
                  Botao Hao and
                  Bowen Baker and
                  Brandon Houghton and
                  Brandon McKinzie and
                  Brydon Eastman and
                  Camillo Lugaresi and
                  Cary Bassin and
                  Cary Hudson and
                  Chak Ming Li and
                  Charles de Bourcy and
                  Chelsea Voss and
                  Chen Shen and
                  Chong Zhang and
                  Chris Koch and
                  Chris Orsinger and
                  Christopher Hesse and
                  Claudia Fischer and
                  Clive Chan and
                  Dan Roberts and
                  Daniel Kappler and
                  Daniel Levy and
                  Daniel Selsam and
                  David Dohan and
                  David Farhi and
                  David Mely and
                  David Robinson and
                  Dimitris Tsipras and
                  Doug Li and
                  Dragos Oprica and
                  Eben Freeman and
                  Eddie Zhang and
                  Edmund Wong and
                  Elizabeth Proehl and
                  Enoch Cheung and
                  Eric Mitchell and
                  Eric Wallace and
                  Erik Ritter and
                  Evan Mays and
                  Fan Wang and
                  Felipe Petroski Such and
                  Filippo Raso and
                  Florencia Leoni and
                  Foivos Tsimpourlas and
                  Francis Song and
                  Fred von Lohmann and
                  Freddie Sulit and
                  Geoff Salmon and
                  Giambattista Parascandolo and
                  Gildas Chabot and
                  Grace Zhao and
                  Greg Brockman and
                  Guillaume Leclerc and
                  Hadi Salman and
                  Haiming Bao and
                  Hao Sheng and
                  Hart Andrin and
                  Hessam Bagherinezhad and
                  Hongyu Ren and
                  Hunter Lightman and
                  Hyung Won Chung and
                  Ian Kivlichan and
                  Ian O'Connell and
                  Ian Osband and
                  Ignasi Clavera Gilaberte and
                  Ilge Akkaya},
  title        = {OpenAI o1 System Card},
  journal      = {CoRR},
  volume       = {abs/2412.16720},
  year         = {2024}
}

@article{Xie2025InterleavedReasoningRL,
  author       = {Roy Xie and
                  David Qiu and
                  Deepak Gopinath and
                  Dong Lin and
                  Yanchao Sun and
                  Chong Wang and
                  Saloni Potdar and
                  Bhuwan Dhingra},
  title        = {Interleaved Reasoning for Large Language Models via Reinforcement
                  Learning},
  journal      = {CoRR},
  volume       = {abs/2505.19640},
  year         = {2025}
}

@article{jiang2025coderlplus,
  author       = {Xue Jiang and
                  Yihong Dong and
                  Mengyang Liu and
                  Hongyi Deng and
                  Tian Wang and
                  Yongding Tao and
                  Rongyu Cao and
                  Binhua Li and
                  Zhi Jin and
                  Wenpin Jiao and
                  Fei Huang and
                  Yongbin Li and
                  Ge Li},
  title        = {CodeRL+: Improving Code Generation via Reinforcement with Execution
                  Semantics Alignment},
  journal      = {CoRR},
  volume       = {abs/2510.18471},
  year         = {2025}
}

@inproceedings{kojima2022zeroshot,
  author       = {Takeshi Kojima and
                  Shixiang Shane Gu and
                  Machel Reid and
                  Yutaka Matsuo and
                  Yusuke Iwasawa},
  title        = {Large Language Models are Zero-Shot Reasoners},
  booktitle    = {NeurIPS},
  year         = {2022}
}

@inproceedings{wang2022selfconsistency,
  author       = {Xuezhi Wang and
                  Jason Wei and
                  Dale Schuurmans and
                  Quoc V. Le and
                  Ed H. Chi and
                  Sharan Narang and
                  Aakanksha Chowdhery and
                  Denny Zhou},
  title        = {Self-Consistency Improves Chain of Thought Reasoning in Language Models},
  booktitle    = {{ICLR}},
  publisher    = {OpenReview.net},
  year         = {2023}
}

@inproceedings{zhou2022leasttomost,
  author       = {Denny Zhou and
                  Nathanael Sch{\"{a}}rli and
                  Le Hou and
                  Jason Wei and
                  Nathan Scales and
                  Xuezhi Wang and
                  Dale Schuurmans and
                  Claire Cui and
                  Olivier Bousquet and
                  Quoc V. Le and
                  Ed H. Chi},
  title        = {Least-to-Most Prompting Enables Complex Reasoning in Large Language
                  Models},
  booktitle    = {{ICLR}},
  publisher    = {OpenReview.net},
  year         = {2023}
}

@inproceedings{yao2023treeofthoughts,
  author       = {Shunyu Yao and
                  Dian Yu and
                  Jeffrey Zhao and
                  Izhak Shafran and
                  Tom Griffiths and
                  Yuan Cao and
                  Karthik Narasimhan},
  title        = {Tree of Thoughts: Deliberate Problem Solving with Large Language Models},
  booktitle    = {NeurIPS},
  year         = {2023}
}

@misc{liang2025plantain,
      title={Plantain: Plan-Answer Interleaved Reasoning}, 
      author={Anthony Liang and Jonathan Berant and Adam Fisch and Abhimanyu Goyal and Kalpesh Krishna and Jacob Eisenstein},
      year={2025},
      eprint={2512.03176},
      archivePrefix={arXiv},
      primaryClass={cs.LG},
      url={https://arxiv.org/abs/2512.03176}, 
}

@article{Kimik2,
  author       = {Yifan Bai and
                  Yiping Bao and
                  Guanduo Chen and
                  Jiahao Chen and
                  Ningxin Chen and
                  Ruijue Chen and
                  Yanru Chen and
                  Yuankun Chen and
                  Yutian Chen and
                  Zhuofu Chen and
                  Jialei Cui and
                  Hao Ding and
                  Mengnan Dong and
                  Angang Du and
                  Chenzhuang Du and
                  Dikang Du and
                  Yulun Du and
                  Yu Fan and
                  Yichen Feng and
                  Kelin Fu and
                  Bofei Gao and
                  Hongcheng Gao and
                  Peizhong Gao and
                  Tong Gao and
                  Xinran Gu and
                  Longyu Guan and
                  Haiqing Guo and
                  Jianhang Guo and
                  Hao Hu and
                  Xiaoru Hao and
                  Tianhong He and
                  Weiran He and
                  Wenyang He and
                  Chao Hong and
                  Yangyang Hu and
                  Zhenxing Hu and
                  Weixiao Huang and
                  Zhiqi Huang and
                  Zihao Huang and
                  Tao Jiang and
                  Zhejun Jiang and
                  Xinyi Jin and
                  Yongsheng Kang and
                  Guokun Lai and
                  Cheng Li and
                  Fang Li and
                  Haoyang Li and
                  Ming Li and
                  Wentao Li and
                  Yanhao Li and
                  Yiwei Li and
                  Zhaowei Li and
                  Zheming Li and
                  Hongzhan Lin and
                  Xiaohan Lin and
                  Zongyu Lin and
                  Chengyin Liu and
                  Chenyu Liu and
                  Hongzhang Liu and
                  Jingyuan Liu and
                  Junqi Liu and
                  Liang Liu and
                  Shaowei Liu and
                  T. Y. Liu and
                  Tianwei Liu and
                  Weizhou Liu and
                  Yangyang Liu and
                  Yibo Liu and
                  Yiping Liu and
                  Yue Liu and
                  Zhengying Liu and
                  Enzhe Lu and
                  Lijun Lu and
                  Shengling Ma and
                  Xinyu Ma and
                  Yingwei Ma and
                  Shaoguang Mao and
                  Jie Mei and
                  Xin Men and
                  Yibo Miao and
                  Siyuan Pan and
                  Yebo Peng and
                  Ruoyu Qin and
                  Bowen Qu and
                  Zeyu Shang and
                  Lidong Shi and
                  Shengyuan Shi and
                  Feifan Song and
                  Jianlin Su and
                  Zhengyuan Su and
                  Xinjie Sun and
                  Flood Sung and
                  Heyi Tang and
                  Jiawen Tao and
                  Qifeng Teng and
                  Chensi Wang and
                  Dinglu Wang and
                  Feng Wang and
                  Haiming Wang},
  title        = {Kimi {K2:} Open Agentic Intelligence},
  journal      = {CoRR},
  volume       = {abs/2507.20534},
  year         = {2025}
}

@misc{lozhkov2024starcoder2stackv2,
      title={StarCoder 2 and The Stack v2: The Next Generation}, 
      author={Anton Lozhkov and Raymond Li and Loubna Ben Allal and Federico Cassano and Joel Lamy-Poirier and Nouamane Tazi and Ao Tang and Dmytro Pykhtar and Jiawei Liu and Yuxiang Wei and Tianyang Liu and Max Tian and Denis Kocetkov and Arthur Zucker and Younes Belkada and Zijian Wang and Qian Liu and Dmitry Abulkhanov and Indraneil Paul and Zhuang Li and Wen-Ding Li and Megan Risdal and Jia Li and Jian Zhu and Terry Yue Zhuo and Evgenii Zheltonozhskii and Nii Osae Osae Dade and Wenhao Yu and Lucas Krauß and Naman Jain and Yixuan Su and Xuanli He and Manan Dey and Edoardo Abati and Yekun Chai and Niklas Muennighoff and Xiangru Tang and Muhtasham Oblokulov and Christopher Akiki and Marc Marone and Chenghao Mou and Mayank Mishra and Alex Gu and Binyuan Hui and Tri Dao and Armel Zebaze and Olivier Dehaene and Nicolas Patry and Canwen Xu and Julian McAuley and Han Hu and Torsten Scholak and Sebastien Paquet and Jennifer Robinson and Carolyn Jane Anderson and Nicolas Chapados and Mostofa Patwary and Nima Tajbakhsh and Yacine Jernite and Carlos Muñoz Ferrandis and Lingming Zhang and Sean Hughes and Thomas Wolf and Arjun Guha and Leandro von Werra and Harm de Vries},
      year={2024},
      eprint={2402.19173},
      archivePrefix={arXiv},
      primaryClass={cs.SE},
      url={https://arxiv.org/abs/2402.19173}, 
}

@article{mbpp,
  author       = {Jacob Austin and
                  Augustus Odena and
                  Maxwell I. Nye and
                  Maarten Bosma and
                  Henryk Michalewski and
                  David Dohan and
                  Ellen Jiang and
                  Carrie J. Cai and
                  Michael Terry and
                  Quoc V. Le and
                  Charles Sutton},
  title        = {Program Synthesis with Large Language Models},
  journal      = {CoRR},
  volume       = {abs/2108.07732},
  year         = {2021}
}

@inproceedings{lora,
  author       = {Edward J. Hu and
                  Yelong Shen and
                  Phillip Wallis and
                  Zeyuan Allen{-}Zhu and
                  Yuanzhi Li and
                  Shean Wang and
                  Lu Wang and
                  Weizhu Chen},
  title        = {LoRA: Low-Rank Adaptation of Large Language Models},
  booktitle    = {{ICLR}},
  publisher    = {OpenReview.net},
  year         = {2022}
}

@article{codeio,
  author       = {Junlong Li and
                  Daya Guo and
                  Dejian Yang and
                  Runxin Xu and
                  Yu Wu and
                  Junxian He},
  title        = {CodeI/O: Condensing Reasoning Patterns via Code Input-Output Prediction},
  journal      = {CoRR},
  volume       = {abs/2502.07316},
  year         = {2025}
}

@article{guo2025deepseek,
  title={{DeepSeek-R1} incentivizes reasoning in {LLMs} through reinforcement learning},
  author={Guo, Daya and Yang, Dejian and Zhang, Haowei and Song, Junxiao and Zhang, Ruoyu and Xu, Runxin and Zhu, Qihao and Ma, Shirong and Wang, Peiyi and Bi, Xiao and Zhang, Xiaokang and Yu, Xingkai and Wu, Yu and Wu, Z. F. and Gou, Zhibin and Shao, Zhihong and Li, Zhuoshu and Gao, Ziyi and Liu, Aixin and Xue, Bing and Wang, Bingxuan and Wu, Bochao and Feng, Bei and Lu, Chengda and Zhao, Chenggang and Deng, Chengqi and Zhang, Chenyu and Ruan, Chong and Dai, Damai and Chen, Deli and Ji, Dongjie and Li, Erhang and Lin, Fangyun and Dai, Fucong and Luo, Fuli and Hao, Guangbo and Chen, Guanting and Li, Guowei and Zhang, H. and Bao, Han and Xu, Hanwei and Wang, Haocheng and Ding, Honghui and Xin, Huajian and Gao, Huazuo and Qu, Hui and Li, Hui and Guo, Jianzhong and Li, Jiashi and Wang, Jiawei and Chen, Jingchang and Yuan, Jingyang and Qiu, Junjie and Li, Junlong and Cai, J. L. and Ni, Jiaqi and Liang, Jian and Chen, Jin and Dong, Kai and Hu, Kai and Gao, Kaige and Guan, Kang and Huang, Kexin and Yu, Kuai and Wang, Lean and Zhang, Lecong and Zhao, Liang and Wang, Litong and Zhang, Liyue and Xu, Lei and Xia, Leyi and Zhang, Mingchuan and Zhang, Minghua and Tang, Minghui and Li, Meng and Wang, Miaojun and Li, Mingming and Tian, Ning and Huang, Panpan and Zhang, Peng and Wang, Qiancheng and Chen, Qinyu and Du, Qiushi and Ge, Ruiqi and Zhang, Ruisong and Pan, Ruizhe and Wang, Runji and Chen, R. J. and Jin, R. L. and Chen, Ruyi and Lu, Shanghao and Zhou, Shangyan and Chen, Shanhuang and Ye, Shengfeng and Wang, Shiyu and Yu, Shuiping and Zhou, Shunfeng and Pan, Shuting and Li, S. S.},
  journal={Nature},
  volume={645},
  number={8081},
  pages={633},
  year={2025}
}

@article{wang2025codeboost,
  title={{CodeBoost}: Boosting Code {LLMs} by Squeezing Knowledge from Code Snippets with RL},
  author={Wang, Sijie and Guo, Quanjiang and Zhao, Kai and Zhang, Yawei and Li, Xin and Li, Xiang and Li, Siqi and She, Rui and Yu, Shangshu and Tay, Wee Peng},
  journal={arXiv preprint arXiv:2508.05242},
  year={2025}
}

@article{guo2024deepseek,
  title={DeepSeek-Coder: When the Large Language Model Meets Programming--The Rise of Code Intelligence},
  author={Guo, Daya and Zhu, Qihao and Yang, Dejian and Xie, Zhenda and Dong, Kai and Zhang, Wentao and Chen, Guanting and Bi, Xiao and Wu, Y. and Li, Y. K. and Luo, Fuli and Xiong, Yingfei and Liang, Wenfeng},
  journal={arXiv preprint arXiv:2401.14196},
  year={2024}
}

@article{jain2024livecodebench,
  title={Livecodebench: Holistic and contamination free evaluation of large language models for code},
  author={Jain, Naman and Han, King and Gu, Alex and Li, Wen-Ding and Yan, Fanjia and Zhang, Tianjun and Wang, Sida and Solar-Lezama, Armando and Sen, Koushik and Stoica, Ion},
  journal={arXiv preprint arXiv:2403.07974},
  year={2024}
}

@article{hui2024qwen2,
  title={Qwen2. 5-coder technical report},
  author={Hui, Binyuan and Yang, Jian and Cui, Zeyu and Yang, Jiaxi and Liu, Dayiheng and Zhang, Lei and Liu, Tianyu and Zhang, Jiajun and Yu, Bowen and Dang, Kai and Yang, An and Men, Rui and Huang, Fei and Ren, Xingzhang and Ren, Xuancheng and Zhou, Jingren and Lin, Junyang},
  journal={arXiv preprint arXiv:2409.12186},
  year={2024}
}

@article{sheng2024hybridflow,
  title   = {HybridFlow: A Flexible and Efficient RLHF Framework},
  author  = {Guangming Sheng and Chi Zhang and Zilingfeng Ye and Xibin Wu and Wang Zhang and Ru Zhang and Yanghua Peng and Haibin Lin and Chuan Wu},
  year    = {2024},
  journal = {arXiv preprint arXiv: 2409.19256}
}

@article{wei2022chain,
  title={Chain-of-thought prompting elicits reasoning in large language models},
  author={Wei, Jason and Wang, Xuezhi and Schuurmans, Dale and Bosma, Maarten and Ichter, Brian and Xia, Fei and Chi, Ed H. and Le, Quoc V. and Zhou, Denny},
  journal={Advances in neural information processing systems},
  volume={35},
  pages={24824--24837},
  year={2022}
}

@article{DeepSeekMath,
  author       = {Zhihong Shao and
                  Peiyi Wang and
                  Qihao Zhu and
                  Runxin Xu and
                  Junxiao Song and
                  Mingchuan Zhang and
                  Y. K. Li and
                  Y. Wu and
                  Daya Guo},
  title        = {DeepSeekMath: Pushing the Limits of Mathematical Reasoning in Open
                  Language Models},
  journal      = {CoRR},
  volume       = {abs/2402.03300},
  year         = {2024}
}

@article{PPO,
  title={Proximal policy optimization algorithms},
  author={Schulman, John and Wolski, Filip and Dhariwal, Prafulla and Radford, Alec and Klimov, Oleg},
  journal={arXiv preprint arXiv:1707.06347},
  year={2017}
}

@article{chen2021evaluating,
  title={Evaluating large language models trained on code},
  author={Chen, Mark and Tworek, Jerry and Jun, Heewoo and Yuan, Qiming and de Oliveira Pinto, Henrique Pondé and Kaplan, Jared and Edwards, Harri and Burda, Yuri and Joseph, Nicholas and Brockman, Greg and Ray, Alex and Puri, Raul and Krueger, Gretchen and Petrov, Michael and Khlaaf, Heidy and Sastry, Girish and Mishkin, Pamela and Chan, Brooke and Gray, Scott and Ryder, Nick and Pavlov, Mikhail and Power, Alethea and Kaiser, Lukasz and Bavarian, Mohammad and Winter, Clemens and Tillet, Philippe and Such, Felipe Petroski and Cummings, Dave and Plappert, Matthias and Chantzis, Fotios and Barnes, Elizabeth and Herbert-Voss, Ariel and Guss, William Hebgen and Nichol, Alex and Paino, Alex and Tezak, Nikolas and Tang, Jie and Babuschkin, Igor and Balaji, Suchir and Jain, Shantanu and Saunders, William and Hesse, Christopher and Carr, Andrew N. and Leike, Jan and Achiam, Joshua and Misra, Vedant and Morikawa, Evan and Radford, Alec and Knight, Matthew and Brundage, Miles and Murati, Mira and Mayer, Katie and Welinder, Peter and McGrew, Bob and Amodei, Dario and McCandlish, Sam and Sutskever, Ilya and Zaremba, Wojciech},
  journal={arXiv preprint arXiv:2107.03374},
  year={2021}
}

@article{gemini2.5,
  author  = {Comanici, Gheorghe and Bieber, Eric and Schaekermann, Mike and Pasupat, Ice and Sachdeva, Noveen and Dhillon, Inderjit S. and Blistein, Marcel and Ram, Ori and Zhang, Dan and Rosen, Evan and Marris, Luke and Petulla, Sam and Gaffney, Colin and Aharoni, Asaf and Lintz, Nathan and Pais, Tiago Cardal and Jacobsson, Henrik and Szpektor, Idan and Jiang, Nan-Jiang and Haridasan, Krishna and Omran, Ahmed and Saunshi, Nikunj and Bahri, Dara and Mishra, Gaurav and Chu, Eric and Boyd, Toby and Hekman, Brad and Parisi, Aaron and Zhang, Chaoyi and Kawintiranon, Kornraphop and Bedrax-Weiss, Tania and Wang, Oliver and Xu, Ya and Purkiss, Ollie and Mendlovic, Uri and Deutel, Ilaï and Nguyen, Nam and Langley, Adam and Korn, Flip and Rossazza, Lucia and Ramé, Alexandre and Waghmare, Sagar and Miller, Helen and Byrd, Nathan and Sheshan, Ashrith and Bhardwaj, Sangnie and Janus, Pawel and Rissa, Tero and Horgan, Dan and Silver, Sharon and Wahid, Ayzaan and Brin, Sergey and Raimond, Yves and Kloboves, Klemen and Wang, Cindy and Gundavarapu, Nitesh Bharadwaj and Shumailov, Ilia and Wang, Bo and Pajarskas, Mantas and Heyward, Joe and Nikoltchev, Martin and Kula, Maciej and Zhou, Hao and Garrett, Zachary and Kafle, Sushant and Arik, Sercan and Goel, Ankita and Yang, Mingyao and Park, Jiho and Kojima, Koji and Mahmoudieh, Parsa and Kavukcuoglu, Koray and Chen, Grace and Fritz, Doug and Bulyenov, Anton and Roy, Sudeshna and Paparas, Dimitris and Shemtov, Hadar and Chen, Bo-Juen and Strudel, Robin and Reitter, David and Roy, Aurko and Vlasov, Andrey and Ryu, Changwan and Leichner, Chas and Yang, Haichuan and Mariet, Zelda and Vnukov, Denis and Sohn, Tim and Stuart, Amy and Liang, Wei and Chen, Minmin and Rawlani, Praynaa and Koh, Christy and Co-Reyes, JD and Lai, Guangda and Banzal, Praseem and Vytiniotis, Dimitrios and Mei, Jieru and Cai, Mu},
  title   = {Gemini 2.5: Pushing the frontier with advanced reasoning, multimodality, long context, and next-generation agentic capabilities},
  journal = {arXiv preprint arXiv:2507.06261},
  year    = {2025}
}

@misc{openr1,
    title = {Open R1: A fully open reproduction of DeepSeek-R1},
    url = {https://github.com/huggingface/open-r1 },
    author = {{Hugging Face}},
    month = {January},
    year = {2025}
}

@article{ocrqwen,
      title={OpenCodeReasoning: Advancing Data Distillation for Competitive Coding}, 
      author={Wasi Uddin Ahmad and Sean Narenthiran and Somshubra Majumdar and Aleksander Ficek and Siddhartha Jain and Jocelyn Huang and Vahid Noroozi and Boris Ginsburg},
      journal={arXiv preprint arXiv:2504.01943},	
      year={2025},
}

@article{skywork,
  title={Skywork Open Reasoner 1 Technical Report},
  author={Jujie He and Jiacai Liu and Chris Yuhao Liu and Rui Yan and Chaojie Wang and Peng Cheng and Xiaoyu Zhang and Fuxiang Zhang and Jiacheng Xu and Wei Shen and Siyuan Li and Liang Zeng and Tianwen Wei and Cheng Cheng and Bo An and Yang Liu and Yahui Zhou},
  journal={arXiv preprint arXiv:2505.22312},
  year={2025}
}

@inproceedings{codePRM,
    author = {Qingyao Li and Xinyi Dai and Xiangyang Li and Weinan Zhang and Yasheng Wang and Ruiming Tang and Yong Yu},
    title = {CodePRM: Execution Feedback-enhanced Process Reward Model for Code Generation},
    booktitle = {Findings of the Association for Computational Linguistics: ACL 2025},
    pages = {8169--8182},
    year = {2025}
}

@article{codereasoner,
    author = {Lingxiao Tang and He Ye and Zhongxin Liu and Xiaoxue Ren and Lingfeng Bao},
    title = {CodeReasoner: Enhancing the Code Reasoning Ability with Reinforcement Learning},
    journal = {arXiv preprint arXiv:2507.17548},
    year = {2025}
}

@misc{llama31,
  author       = {{Meta AI}},
  title        = {Introducing Llama 3.1: Our most capable models to date},
  year         = {2024},
  month        = {jul},
  day          = {23},
  howpublished = {\url{https://ai.meta.com/blog/meta-llama-3-1/ }},
  note         = {Accessed: 2025-10-06}
}

@article{codellama,
    author = {Rozière, Baptiste and Gehring, Jonas and Gloeckle, Fabian and Sootla, Sten and Gat, Itai and Tan, Xiaoqing Ellen and Adi, Yossi and Liu, Jingyu and Sauvestre, Romain and Remez, Tal and Rapin, Jérémy and Kozhevnikov, Artyom and Evtimov, Ivan and Bitton, Joanna and Bhatt, Manish and Ferrer, Cristian Canton and Grattafiori, Aaron and Xiong, Wenhan and Défossez, Alexandre and Copet, Jade and Azhar, Faisal and Touvron, Hugo and Martin, Louis and Usunier, Nicolas and Scialom, Thomas and Synnaeve, Gabriel},
    title = {Code llama: Open foundation models for code},
    journal = {arXiv preprint arXiv:2308.12950},
    year = {2023}
}

@article{leetcode,
    author = {Yunhui Xia and Wei Shen and Yan Wang and Jason Klein Liu and Huifeng Sun and Siyue Wu and Jian Hu and Xiaolong Xu},
    title = {Leetcodedataset: A temporal dataset for robust evaluation and efficient training of code llms},
    journal = {arXiv preprint arXiv:2504.14655},
    year = {2025}
}

@article{DBLP:journals/tosem/DongJJL24,
  author       = {Yihong Dong and
                  Xue Jiang and
                  Zhi Jin and
                  Ge Li},
  title        = {Self-Collaboration Code Generation via ChatGPT},
  journal      = {{ACM} Trans. Softw. Eng. Methodol.},
  volume       = {33},
  number       = {7},
  pages        = {189:1--189:38},
  year         = {2024}
}

@inproceedings{DBLP:conf/acl/DongJLJGYL24,
  author       = {Yihong Dong and
                  Xue Jiang and
                  Huanyu Liu and
                  Zhi Jin and
                  Bin Gu and
                  Mengfei Yang and
                  Ge Li},
  title        = {Generalization or Memorization: Data Contamination and Trustworthy
                  Evaluation for Large Language Models},
  booktitle    = {{ACL} (Findings)},
  series       = {Findings of {ACL}},
  pages        = {12039--12050},
  publisher    = {Association for Computational Linguistics},
  year         = {2024}
}

@article{DBLP:journals/corr/abs-2508-00083,
  author       = {Yihong Dong and
                  Xue Jiang and
                  Jiaru Qian and
                  Tian Wang and
                  Kechi Zhang and
                  Zhi Jin and
                  Ge Li},
  title        = {A Survey on Code Generation with LLM-based Agents},
  journal      = {CoRR},
  volume       = {abs/2508.00083},
  year         = {2025}
}

@inproceedings{
pfau2024lets,
title={Let{\textquoteright}s Think Dot by Dot: Hidden computation in transformer language models},
author={Jacob Pfau and William Merrill and Samuel R. Bowman},
booktitle={First Conference on Language Modeling},
year={2024},
url={https://openreview.net/forum?id=NikbrdtYvG}
}

@inproceedings{
goyal2024think,
title={Think before you speak: Training Language Models With Pause Tokens},
author={Sachin Goyal and Ziwei Ji and Ankit Singh Rawat and Aditya Krishna Menon and Sanjiv Kumar and Vaishnavh Nagarajan},
booktitle={The Twelfth International Conference on Learning Representations},
year={2024},
url={https://openreview.net/forum?id=ph04CRkPdC}
}

@article{DBLP:journals/corr/abs-2601-13240,
  author       = {Xue Jiang and
                  Jiaru Qian and
                  Xianjie Shi and
                  Chenjie Li and
                  Hao Zhu and
                  Ziyu Wang and
                  Jielun Zhang and
                  Zheyu Zhao and
                  Kechi Zhang and
                  Jia Li and
                  Wenpin Jiao and
                  Zhi Jin and
                  Ge Li and
                  Yihong Dong},
  title        = {{KOCO-BENCH:} Can Large Language Models Leverage Domain Knowledge
                  in Software Development?},
  journal      = {CoRR},
  volume       = {abs/2601.13240},
  year         = {2026}
}

@article{DBLP:journals/corr/abs-2508-00222,
  author       = {Yihong Dong and
                  Xue Jiang and
                  Yongding Tao and
                  Huanyu Liu and
                  Kechi Zhang and
                  Lili Mou and
                  Rongyu Cao and
                  Yingwei Ma and
                  Jue Chen and
                  Binhua Li and
                  Zhi Jin and
                  Fei Huang and
                  Yongbin Li and
                  Ge Li},
  title        = {{RL-PLUS:} Countering Capability Boundary Collapse of LLMs in Reinforcement
                  Learning with Hybrid-policy Optimization},
  journal      = {CoRR},
  volume       = {abs/2508.00222},
  year         = {2025}
}

@misc{aime24,
  author       = {{Mathematical Association of America}},
  title        = {American Invitational Mathematics Examination (AIME) 2024},
  howpublished = {\url{https://maa.org/maa-invitational-competitions/}},
  year         = {2024},
}

@misc{aime25,
  author       = {{Mathematical Association of America}},
  title        = {American Invitational Mathematics Examination (AIME) 2025},
  howpublished = {\url{https://maa.org/maa-invitational-competitions/}},
  year         = {2025},
}

@misc{hmmt25,
  title = {MathArena: Evaluating LLMs on Uncontaminated Math Competitions},
  author = {Mislav Balunović and Jasper Dekoninck and Ivo Petrov and Nikola Jovanović and Martin Vechev},
  copyright = {MIT},
  url = {https://matharena.ai/},
  publisher = {SRI Lab, ETH Zurich},
  month = feb,
  year = {2025},
}
\bibliographystyle{preprint}

\newpage
\appendix

\section{Token Cost Breakdown}
\label{appendix:token_breakdown}

Table~\ref{tab:token_breakdown} provides a detailed breakdown of reasoning token costs. For GRPO and CoT, the token cost consists entirely of upfront thinking tokens. For \ourapproach, we separately report the upfront thinking length and the \texttt{<thinkanywhere>} block length. The upfront thinking phase of \ourapproach is substantially shorter than that of GRPO and CoT across benchmarks, and the additional \texttt{<thinkanywhere>} tokens are modest in comparison, resulting in a net reduction in total reasoning token usage.

\begin{table}[h]
\centering
\caption{Breakdown of reasoning token costs. For \ourapproach, the two numbers denote upfront thinking length + \texttt{<thinkanywhere>} block length.}
\label{tab:token_breakdown}
\begin{tabular}{lccc}
\toprule
\textbf{Method} & \textbf{HumanEval} & \textbf{MBPP} & \textbf{LeetCode} \\
\midrule
GRPO & 309.4 & 325.2 & 440.7 \\
CoT & 348.8 & 372.0 & 577.0 \\
\ourapproach & 215.6 + 22.5 & 183.2 + 23.2 & 283.0 + 22.9 \\
\bottomrule
\end{tabular}
\end{table}

\section{Thinking Block Statistics Across Training Stages}
\label{appendix:thinking_stats}

To clarify the respective contributions of cold-start SFT and RLVR, we analyze the average frequency (Avg.Freq) and average length (Avg.Len) of \texttt{<thinkanywhere>} blocks across different training stages, as shown in Table~\ref{tab:thinking_stats}.

The base model never invokes thinking blocks during code generation. Even with explicit prompting (\ourapproach Prompting), the model produces very few \texttt{<thinkanywhere>} blocks (frequency near zero) with abnormally long lengths, indicating that prompting alone cannot reliably elicit \ourapproach behavior and that this capability is unlikely to originate from pre-training. After cold-start SFT, the model generates \texttt{<thinkanywhere>} blocks at a normal frequency and length, demonstrating that SFT successfully teaches the model to imitate the \ourapproach reasoning pattern and establishes a solid foundation for subsequent RL training. After RL training, the frequency and length of \texttt{<thinkanywhere>} blocks decrease slightly compared to the SFT stage, while pass@1 improves substantially (as shown in Table~\ref{tab:ablation}). This indicates that RL does not simply increase the number of thinking tokens; rather, it refines the model's ability to invoke reasoning on demand, enabling more concise and targeted deliberation at positions where it is truly needed. This is fully consistent with the design goal of \ourapproach.

\begin{table}[h]
\centering
\caption{Average frequency and length of \texttt{<thinkanywhere>} blocks across training stages.}
\label{tab:thinking_stats}
\begin{tabular}{llcc}
\toprule
\textbf{Dataset} & \textbf{Model} & \textbf{Avg.Freq} & \textbf{Avg.Len} \\
\midrule
\multirow{4}{*}{HumanEval} & Base Model & 0 & 0 \\
& \ourapproach (Prompting) & 0.24 & 113.5 \\
& \ourapproach (SFT) & 6.69 & 31.9 \\
& \ourapproach (Ours) & 6.15 & 22.5 \\
\midrule
\multirow{4}{*}{MBPP} & Base Model & 0 & 0 \\
& \ourapproach (Prompting) & 0.53 & 66.4 \\
& \ourapproach (SFT) & 5.76 & 33.4 \\
& \ourapproach (Ours) & 5.24 & 23.2 \\
\midrule
\multirow{4}{*}{LeetCode} & Base Model & 0 & 0 \\
& \ourapproach (Prompting) & 0.31 & 219.7 \\
& \ourapproach (SFT) & 11.28 & 34.5 \\
& \ourapproach (Ours) & 11.26 & 22.9 \\
\bottomrule
\end{tabular}
\end{table}

\end{document}